\documentclass[
superscriptaddress,
reprint,
longbibliography,
floatfix,
amsmath,amssymb,
aps,
prl,
]{revtex4-2}

\usepackage[dvipsnames]{xcolor}
\usepackage{graphicx}
\usepackage{bm}
\usepackage{hyperref}
\usepackage{orcidlink}
\hypersetup{colorlinks=true,urlcolor=blue,citecolor=blue}

\usepackage{changes}

\begin{document}

\title{Unoccupied spin texture of ordered Bi films on vicinal Ge substrates}

\author{F. Goto\,\orcidlink{0000-0002-0948-5951}}
\email{francesco.goto@polimi.it}
\affiliation{Dipartimento di Fisica, Politecnico di Milano, Piazza Leonardo da Vinci 32, 20133 Milano, Italy}

\author{G. Bussetti\,\orcidlink{0000-0001-8556-8014}}
\affiliation{Dipartimento di Fisica, Politecnico di Milano, Piazza Leonardo da Vinci 32, 20133 Milano, Italy}

\author{L. Nicola\"i\,\orcidlink{0000-0002-7731-2673}}
\affiliation{New Technologies - Research Centre, University of West Bohemia in Pilsen, Teslova 9a, 301 00 Pilsen, Czech Republic}

\author{A. Pulkkinen\,\orcidlink{0000-0002-4339-6928}}
\affiliation{New Technologies - Research Centre, University of West Bohemia in Pilsen, Teslova 9a, 301 00 Pilsen, Czech Republic}

\author{C. Zucchetti\,\orcidlink{0000-0002-0428-5184}}
\affiliation{Dipartimento di Fisica, Politecnico di Milano, Piazza Leonardo da Vinci 32, 20133 Milano, Italy}

\author{F. Bottegoni\,\orcidlink{0000-0002-3108-0563}}
\affiliation{Dipartimento di Fisica, Politecnico di Milano, Piazza Leonardo da Vinci 32, 20133 Milano, Italy}

\author{M. Finazzi\,\orcidlink{0000-0002-9197-3654}}
\affiliation{Dipartimento di Fisica, Politecnico di Milano, Piazza Leonardo da Vinci 32, 20133 Milano, Italy}

\author{L. Du\`o\,\orcidlink{0000-0002-5523-9305}}
\affiliation{Dipartimento di Fisica, Politecnico di Milano, Piazza Leonardo da Vinci 32, 20133 Milano, Italy}

\author{F. Ciccacci\,\orcidlink{0000-0002-1131-4748}}
\affiliation{Dipartimento di Fisica, Politecnico di Milano, Piazza Leonardo da Vinci 32, 20133 Milano, Italy}

\author{J. Min\'ar\,\orcidlink{0000-0001-9735-8479}}
\affiliation{New Technologies - Research Centre, University of West Bohemia in Pilsen, Teslova 9a, 301 00 Pilsen, Czech Republic}

\author{A. Calloni\,\orcidlink{0000-0002-7389-2703}}
\email{alberto.calloni@polimi.it}
\affiliation{Dipartimento di Fisica, Politecnico di Milano, Piazza Leonardo da Vinci 32, 20133 Milano, Italy}

\date{\today}

\begin{abstract}
Vicinal surfaces offer a compelling testing ground in surface science: breaking the high symmetry of a singular surface introduces periodic step arrays that can modify surface states on the terraces and influence epitaxial ordering; the interplay between these effects governs morphology and electronic structure, especially in heavy, spin--orbit-coupled materials. We combine low-energy electron diffraction (LEED) and spin-polarized inverse photoemission (SP-IPES) with \textit{ab initio} calculations for the Bi(111) reference, for the first time on a true high-index vicinal bismuth surface. On the ordered Bi$(9\,9\,11)$ vicinal film selected by Ge(223), the measured unoccupied-state spin-sign sequence matches the calculated initial-state spectral function of Bi(111) terraces. Conversely, the pseudo-cubic Bi domains retained on Ge(443) and Ge(331) exhibit a characteristic spin-polarized surface-state response referenced to the macroscopic vicinal surface. These results show that the substrate cut selects not only the morphology of the Bi film, but also the crystallographic reference frame of its unoccupied spin texture.
\end{abstract}

\maketitle

Spin--orbit coupling at crystal surfaces can produce spin-polarized electronic states whose texture is constrained by surface symmetry. Vicinal surfaces provide a controlled way to modify this symmetry: a periodic array of atomic steps introduces a one-dimensional direction and a well-defined terrace width over macroscopic areas. In layered materials, this geometry turns step configurations otherwise occurring as isolated or irregular defects into reproducible structures. At the same time, it defines two inequivalent crystallographic reference frames---the local low-index terraces and the macroscopic surface formed by the step array---raising the question of which frame governs the electronic states. On vicinal noble-metal surfaces, changes of the electronic reference frame have been attributed to the competition between terrace confinement and electronic effective mass \cite{Ortega2013,Ortega2000}.

Bi provides a particularly suitable test for the influence of periodic steps in a material with strong spin--orbit coupling. Bi(111) hosts metallic spin-polarized surface states \cite{HOFMANN2006,ast2001fermi,Ast2003,Hricovini2019}, while one-dimensional states at Bi step edges have been discussed in connection with higher-order topology \cite{Drozdov2014,Liu2017,Peng2018,Higher-orderBismuth2018,Nayak2019,Wells2009,Bianchi2015,Hofmann2019,Asland2023}. An ordered vicinal Bi(111) film would therefore allow us to investigate how a regular array of steps modifies the established Bi(111) spin texture. Such surfaces are difficult to obtain directly from bulk Bi because of its layered structure, and previous attempts based on Bi deposition on vicinal Si$(h,h,k)$ substrates produced rotationally disordered films \cite{Lukermann2014}. Here we instead use vicinal Ge substrates as epitaxial templates, preparing well-defined flat and high-index surfaces with different miscut directions and angles around [111] \cite{Jeong1999,Tegenkamp2009,Leroy2007,luth2015}.

The growth of ordered hexagonal Bi films on semiconductor substrates is itself nontrivial. At reduced thickness, Bi tends to form a metastable pseudo-cubic phase with a pseudo-square, Bi(110)-type surface lattice, referred to here as PC-Bi \cite{Yaginuma2007,Lukermann2014,Goto2021}.

A second difficulty concerns the interpretation of the measured spin polarization. In spin-resolved photoemission from Bi, final-state and transition-matrix-element interference can generate or strongly reshape the photoelectron polarization, so that the measured signal does not necessarily reproduce the initial-state spin texture \cite{Kimura2010}. Spin-polarized inverse photoemission (SP-IPES), which probes the electronic structure above the Fermi level $E_{\mathrm{F}}$, can likewise depend on transition matrix elements and experimental geometry, particularly when states of mixed orbital symmetry contribute \cite{Wissing2014}.
A comparison between the experimental momentum-dependent spin-sign sequence and the calculated initial-state spin spectral function can therefore help disentangle these contributions. We address these issues by comparing SP-IPES measurements with fully relativistic calculations, focusing on the dominant spin component and its reversal around $\bar{\Gamma}$ above {$E_{\mathrm{F}}$}.

\begin{figure*}[t]
 \centering
 \includegraphics[width=0.8\textwidth]{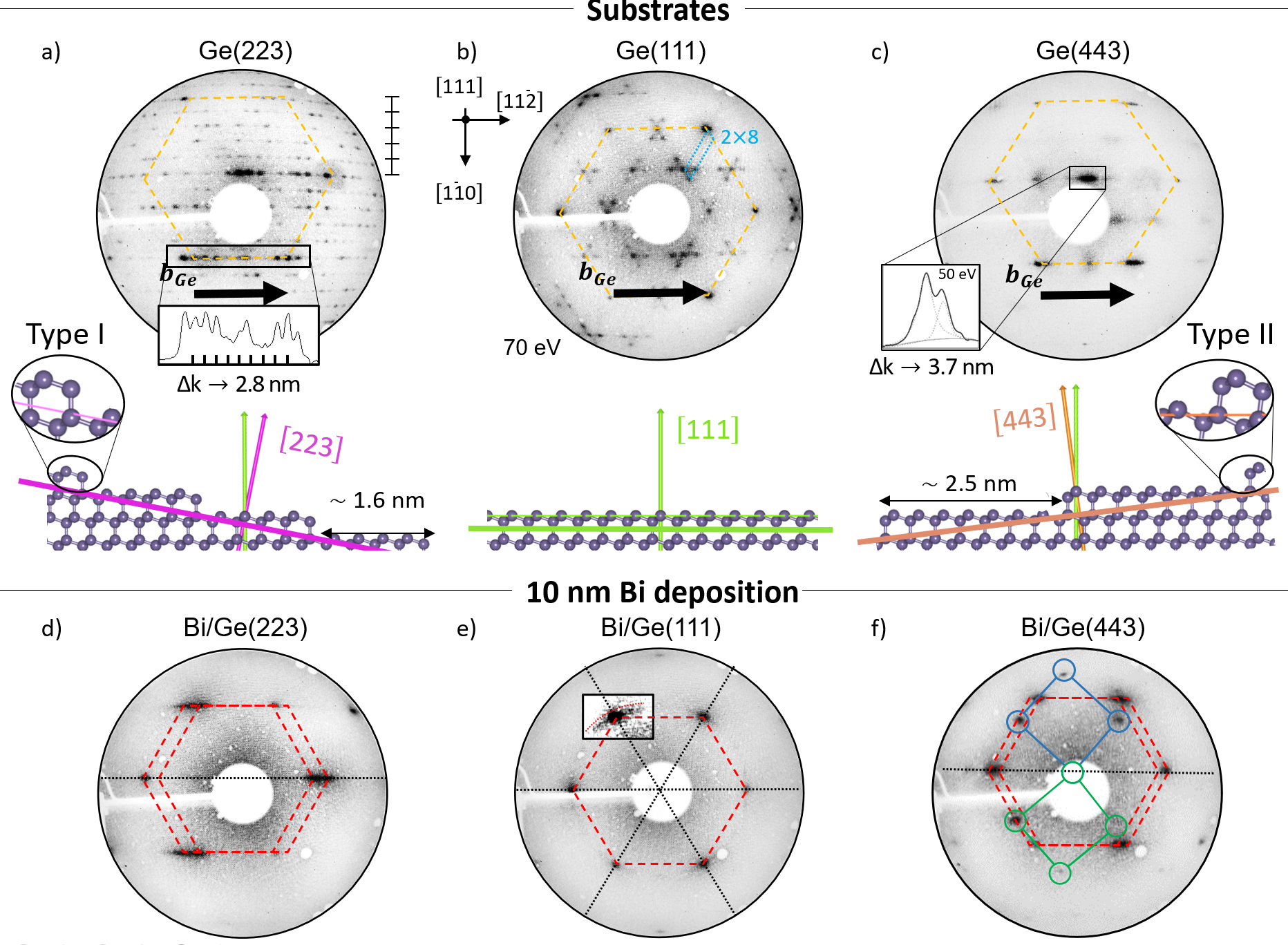}
 \caption{\textbf{Cut-dependent structure of Bi films on Ge.}
 Low-energy electron diffraction (LEED) patterns of clean Ge(223), Ge(111), and Ge(443) (a--c), and LEED after deposition of 10 nm Bi (d--f). The ball-and-stick models below the top-row patterns illustrate ideal bulk-truncated Ge(223), Ge(111), and Ge(443) geometries, with Type I and Type II denoting opposite step-edge configurations. The LEED patterns were acquired at 70 eV; the Ge(443) inset shows the 50-eV profile used to estimate the terrace period. The dashed hexagons mark the Ge and Bi(111) reciprocal lattices; the squares on Bi/Ge(443) mark two mirror-related PC-Bi domains. Flat Ge(111) yields Bi(111) with residual rotational replicas. Ge(443) contains hexagonal Bi(111) spots with reduced splitting, together with PC-Bi diffraction features. Ge(223) yields a dominant vicinal Bi(111) pattern without a resolved PC-Bi contribution.}
 \label{fig:Fig1}
\end{figure*}

Here we compare identically prepared 10-nm Bi films grown on Ge substrates with different vicinal orientations to determine the crystallographic reference frame followed by their unoccupied electronic structure. LEED shows that Ge(223) supports an ordered vicinal Bi(111), consistent with a Bi$(9\,9\,11)$ termination, whereas Ge(443) stabilizes both hexagonal and PC-Bi domains. SP-IPES then determines whether the characteristic surface-state response follows the local terraces or the macroscopic vicinal surface.

Three substrates exhibiting distinct surface reconstructions (Fig.~\ref{fig:Fig1}, top row) are employed. The Ge(443) and Ge(223) spot splittings correspond to average step periods of 3.7 and 2.8 nm, respectively (see Supplemental Material, Sec.~IX~\cite{SupplementalMaterial}). For Ge(223), the measured period exceeds the 1.6-nm value expected for bulk-truncated monoatomic steps and is instead consistent with the presence of biatomic steps, as reported for Ge(223) facets \cite{Wagner2018}. On Ge(443), the residual intensity of the characteristic $c(2\times8)$ reconstruction of Ge(111) suggests the presence of large (111) terraces produced by step bunching.

After deposition of a $\sim10$-nm Bi film at room temperature (in the hexagonal-growth regime established on flat Bi/Ge(111)~\cite{Zucchetti2017,Goto2021}), followed by post-annealing at 400 K (see Supplemental Material, Sec.~I for details \cite{SupplementalMaterial,Olshanetsky1977,Olshanetsky1979,Olshanetsky1981,Allen1957,Brahlek2020,Liu2020,Goto2023,Zucchetti2017,Hirahara2006,Hirahara2015,Takayama2015,Mayer1993,Johnson1997,Berti2014,Finazzi1993,Ciccacci1992,Ciccacci1997,TOGAWA1998,Donath2015}), LEED shows that flat Ge(111) supports epitaxial Bi(111) [Fig.~\ref{fig:Fig1}(e)], satisfying $8a_{Bi}=9a_{Ge}$, with only weak rotational replicas \cite{Hatta2009,Hatta2009LEEED}.
Hexagonal domains are also present on Ge(443) and Ge(223) [Figs.~\ref{fig:Fig1}(f) and \ref{fig:Fig1}(d), respectively]. On Ge(223) [Fig.~\ref{fig:Fig1}(d)], their well-resolved splitting identifies a single vicinal hexagonal registry. On Ge(443) [Fig.~\ref{fig:Fig1}(f)], the reduced splitting of the hexagonal spots indicates a terrace period of approximately 4 nm, whereas no terrace-induced splitting is resolved for the two coexisting mirror-related PC-Bi patterns.

The Ge(443) and Ge(223) surfaces differ not only in terrace period, but also in surface reconstruction and miscut direction, corresponding to two different bulk-truncated step geometries (Type I and Type II in Fig.~\ref{fig:Fig1}). Experimentally, the wider Ge(443) terraces coexist with PC-Bi, whereas the denser Ge(223) step array stabilizes ordered vicinal Bi(111) domains. This comparison already indicates that terrace width alone is insufficient to account for the resulting Bi phase.
\begin{figure}[tbp]
 \centering
 \includegraphics[width=0.99\columnwidth]{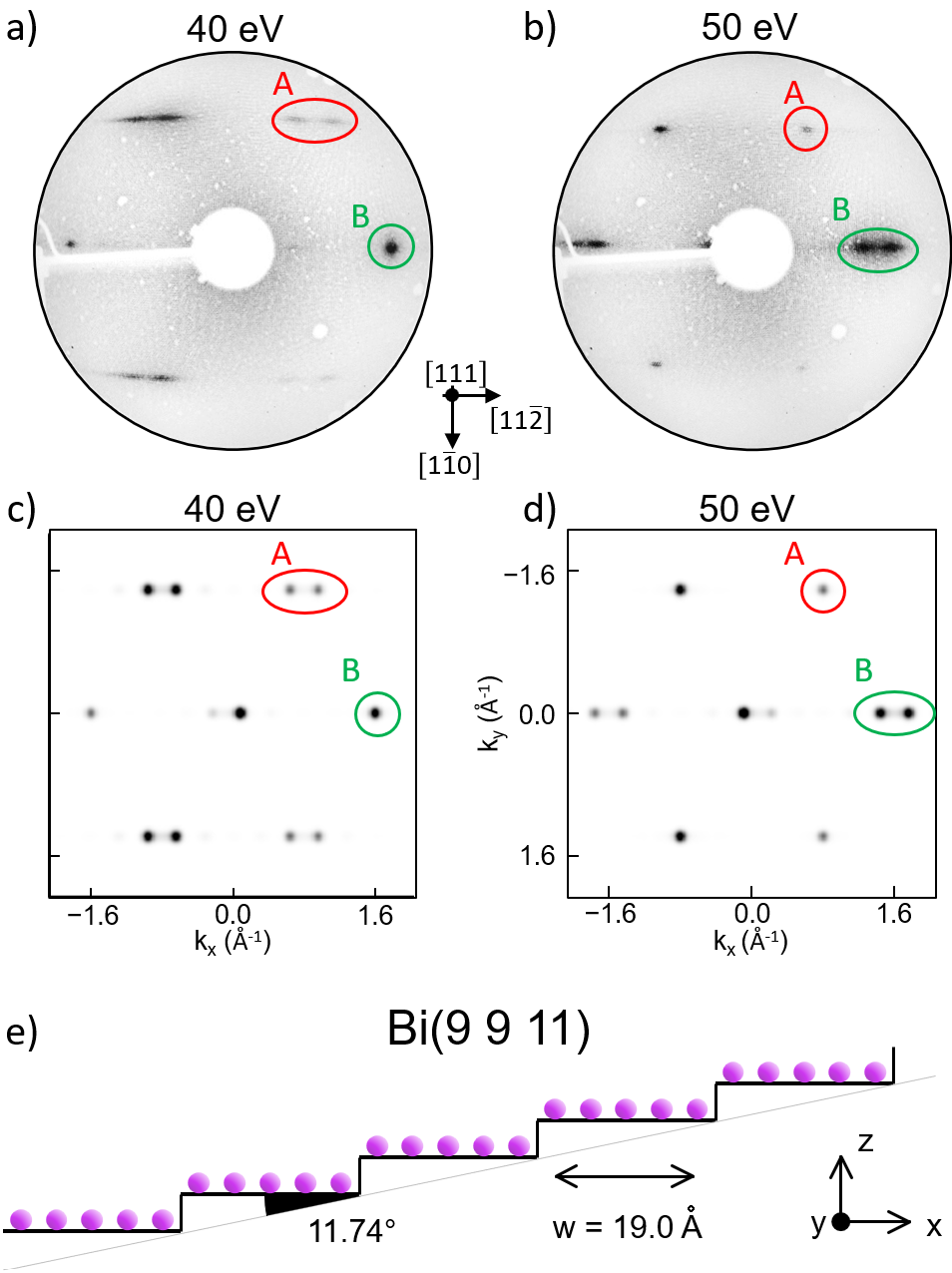}
 \caption{\textbf{Bi$(9\,9\,11)$ films on Ge(223).}
 LEED patterns at 40 and 50 eV in (a) and (b), respectively. The measured splitting corresponds to a Bi terrace period of $2.0\pm0.5$ nm. Panels (c) and (d) show the corresponding reduced kinematic LEED patterns calculated for the mean Bi$(9\,9\,11)$ terrace geometry in panel (e), including the effective inter-terrace phase required to reproduce the relative spot intensities.}
 \label{fig:Fig2}
\end{figure}

Focusing on the Bi film grown on Ge(223), Figs.~\ref{fig:Fig2}(a) and \ref{fig:Fig2}(b) show the peculiar energy dependence of diffraction spots A and B, labelled in Figs.~\ref{fig:Fig2}(a) and \ref{fig:Fig2}(b): their momentum splitting is alternately visible at 40 and 50 eV. Finite domains could broaden or split a diffraction spot, but this alone does not account for the observed alternation between split and unsplit spots, which is consistent with coherent scattering across terraces. The measured splitting over the investigated LEED energies, \mbox{$\Delta k=0.32\pm0.05\,\text{\AA}^{-1}$}, corresponds to a real-space periodicity of $2.0\pm0.5$ nm. Assuming a single Bi bilayer step height, this gives a miscut angle of $\sim11.3^\circ\pm1.5^\circ$, in excellent agreement with both the Ge(223) substrate and the nominal geometry of a Bi$(9\,9\,11)$ vicinal crystal ($11.7^\circ$, 19~\AA). 

The mean Bi$(9\,9\,11)$ terrace geometry fixes the diffraction-splitting scale. The reduced kinematic patterns in Figs.~\ref{fig:Fig2}(c) and \ref{fig:Fig2}(d) also include an effective inter-terrace phase corresponding to a half-row shift along the step edges: this additional phase accounts for the energy-dependent alternation between split and unsplit spots A and B (see Supplemental Material, Sec.~VIII~\cite{SupplementalMaterial}). The diffraction measurements support an ordered Bi$(9\,9\,11)$ average geometry, where the local Bi(111) planes align with the Ge terraces while the average Bi surface remains parallel to the Ge(223) macroscopic surface.

\begin{figure}[tbp]

 \centering
 \includegraphics[width=0.99\columnwidth]{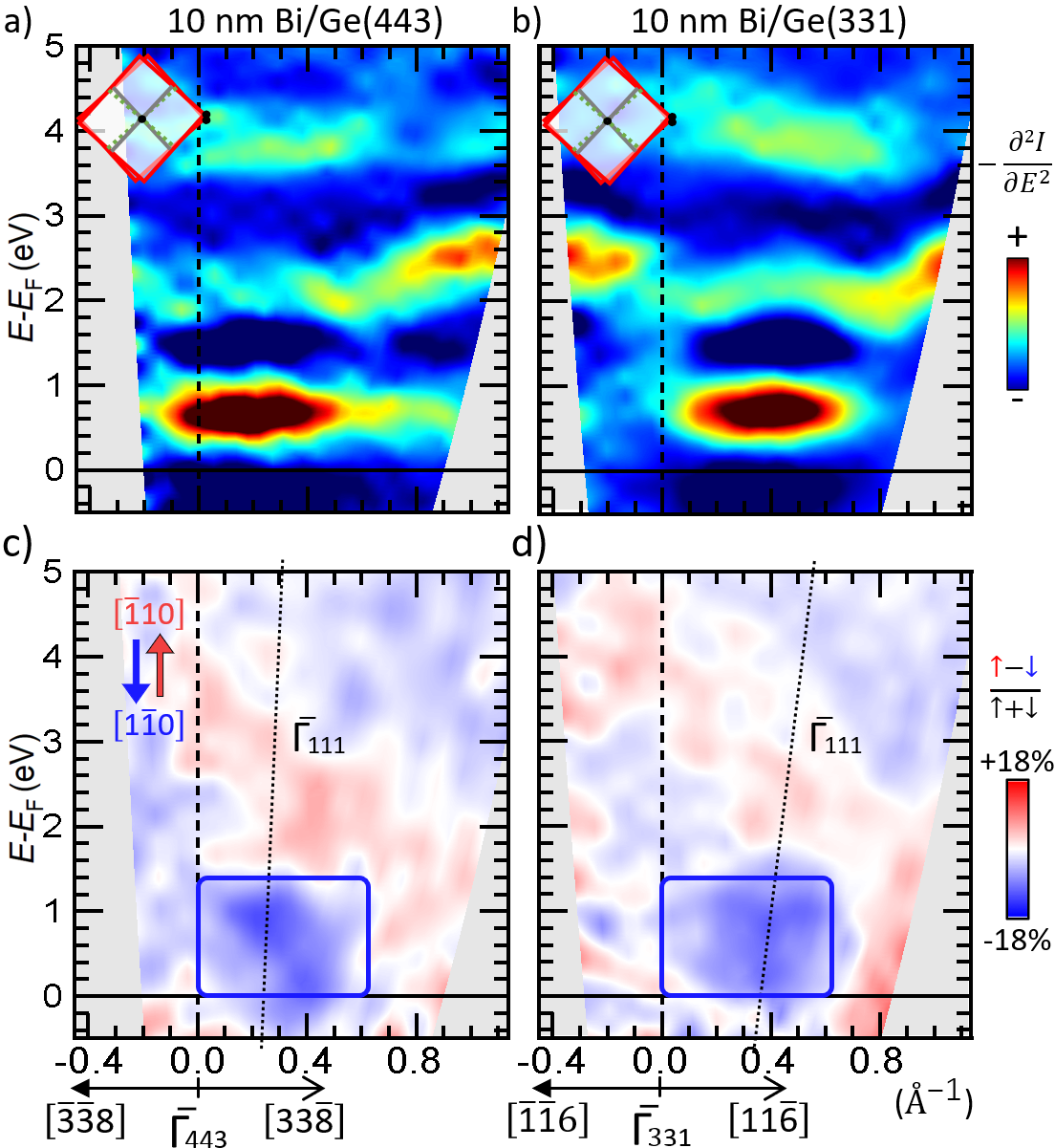}
 \caption{ \textbf{PC-Bi surface-state response in the macroscopic vicinal-surface frame.}
 Second-derivative inverse-photoemission (IPES) intensity in (a) and (b), and spin polarization in (c) and (d), for 10 nm Bi/Ge(443) and Bi/Ge(331). Both films show the same PC-Bi surface-state feature near \mbox{0.7~eV} and a similar spin pattern (blue boxes) when momentum is referenced to the macroscopic vicinal-surface normal (dashed lines). The dotted lines mark the projected $\bar{\Gamma}_{111}$ point. The insets in panels (a) and (b) indicate the Brillouin zones of the selected PC-Bi domains (see Supplemental Material, Secs.~V and VI \cite{SupplementalMaterial}).}
 \label{fig:Fig3}
\end{figure}

We now use SP-IPES to determine whether the measured surface-state spin polarization follows the macroscopic vicinal surface or the local Bi terraces. The reference frame is identified from the momentum about which the polarization reverses, as required by time-reversal symmetry.

\begin{figure*}[t]
 \centering
 \includegraphics[width=0.87\textwidth]{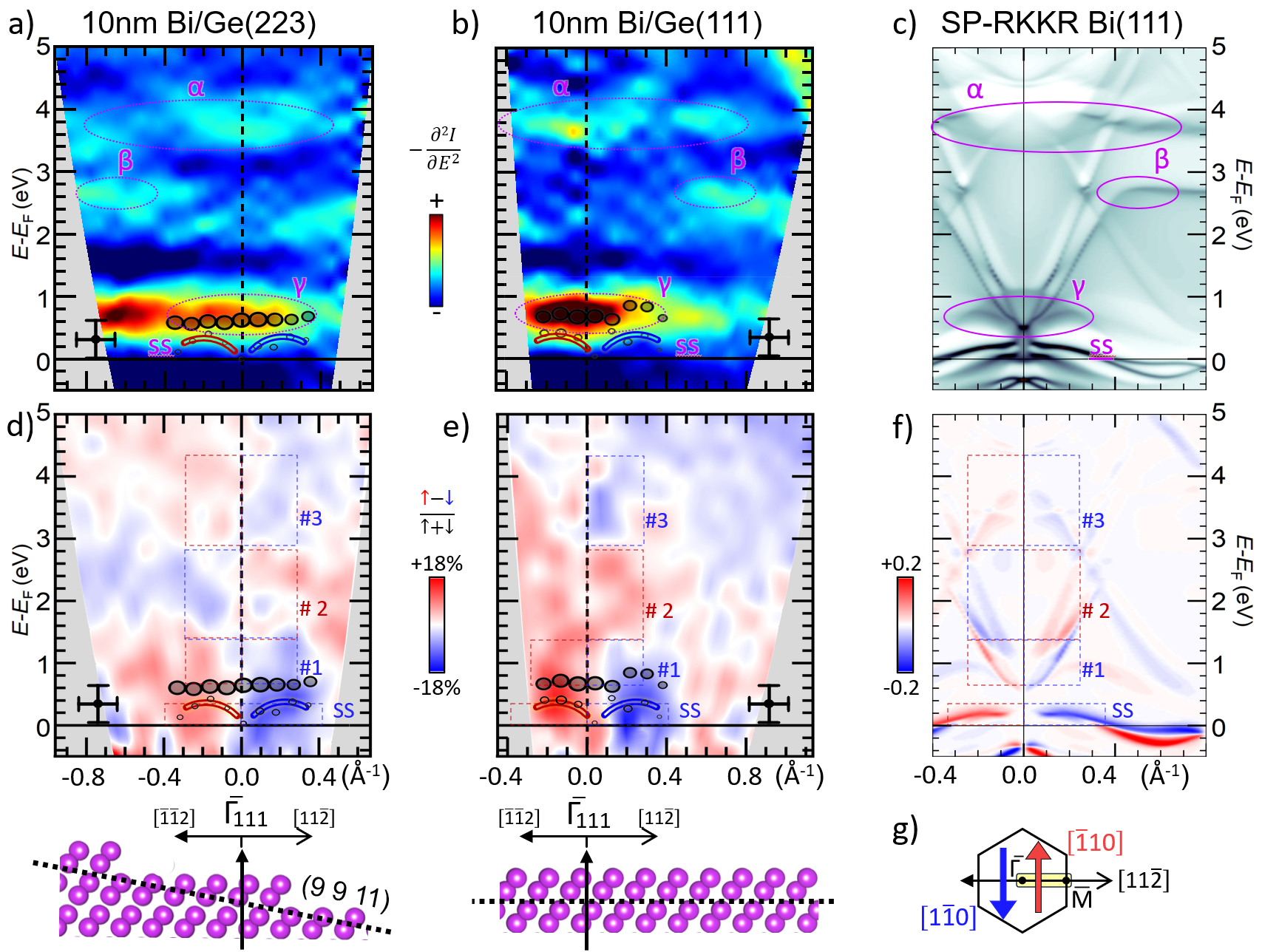}
 \caption{\textbf{Unoccupied Bi(111) spin polarization in the local terrace frame.}
 Second-derivative IPES intensity for 10 nm Bi/Ge(223) (a) and Bi/Ge(111) (b); fully relativistic Korringa--Kohn--Rostoker (KKR) spin-integrated spectral function for Bi(111) (c). The corresponding experimental SP-IPES polarization is shown in (d) and (e), and the calculated Bi(111) spin polarization in (f). On Bi/Ge(223), $k_{111}=0$ is the projected $\bar{\Gamma}_{111}$ of the tilted terrace. The regions labelled \#1--\#3 mark the corresponding sign domains used to compare experiment and the calculated unoccupied-state spin spectral function. Shaded and open circles mark the spin-integrated and spin-resolved fitted components, respectively; their radii scale with the corresponding peak areas. (g) Sketch indicating the measured spin component (red: positive; blue: negative) along $[\bar{1}10]$, perpendicular to the momentum scan $\bar{\Gamma}\bar{M}$.}
 \label{fig:Fig4}
\end{figure*}

On Ge(443), hexagonal and PC-Bi phases coexist. However, the IPES intensity is dominated by the characteristic PC-Bi surface-state feature [Fig.~\ref{fig:Fig3}(a)]. We assign it to the unoccupied continuation of the high-momentum state observed by spin-resolved photoemission \cite{Zucchetti2018}; the corresponding empty-state feature was also detected by SP-IPES and scanning tunneling spectroscopy on PC-Bi islands \cite{Goto2021}. Ultraviolet photoemission (UPS) measurements confirm that Bi(111) states are also present in the occupied electronic structure (see Supplemental Material, Sec.~II \cite{SupplementalMaterial}). The measured spin polarization remains finite at the momentum corresponding to the projected $\bar{\Gamma}_{111}$ of the local Bi(111) terraces (dotted line), whereas it vanishes at the macroscopic surface normal (dashed line), indicating that the dominant PC-Bi surface-state response is referenced to the macroscopic vicinal surface. LEED identifies the PC-Bi domains most clearly for incidence along [111], but this diffraction geometry does not determine the electronic reference frame. The vicinal-referenced response is consistent with a delocalized surface-state regime analogous to that observed on vicinal noble metals \cite{Ortega2000,Ortega2013}.

This interpretation is reinforced by the comparison with Bi films deposited on Ge(331), a vicinal Ge(110) surface that lacks extended (111) terraces but shares the same Type-II step-edge geometry as Ge(443). Bi/Ge(331) exhibits the same PC-Bi domains observed on Ge(443); their LEED spots become clearly resolved for electron-beam incidence along [111], suggesting a common [111]-referenced registry associated with Type-II step edges (see Supplemental Material, Sec.~V \cite{SupplementalMaterial}). More importantly, the two films exhibit the same intense PC-Bi feature near 0.7~eV [Figs.~\ref{fig:Fig3}(a) and \ref{fig:Fig3}(b)] and the same spin-polarization pattern [blue boxes in Figs.~\ref{fig:Fig3}(c) and \ref{fig:Fig3}(d)]. In both cases, the PC-Bi surface-state response is referenced to the macroscopic vicinal plane, whereas the Bi(111) domains present on Ge(443) contribute predominantly to the occupied electronic structure (see Supplemental Material, Secs.~II, V, and VI \cite{SupplementalMaterial}). Since the two phases have different densities of states and transition matrix elements, the relative IPES intensities between PC-Bi and Bi(111) domains cannot be quantitatively compared.

On Ge(223), the ordered vicinal film, consistent with a Bi$(9\,9\,11)$ termination, provides a highly ordered system in which the rotational disorder observed on Bi/Ge(111) is strongly reduced. The momentum scan is performed normal to the step edges, along the local $\bar{\Gamma}\bar{M}$ direction, probing the in-plane spin component transverse to the momentum, \textit{i.e.} parallel to $[\bar{1}10]$ (Fig.~\ref{fig:Fig4}). On flat Bi/Ge(111), $k_{111}=0$ corresponds to normal electron incidence referenced to the macroscopic surface normal, whereas on Bi$(9\,9\,11)$ it is shifted by the miscut angle and coincides with the projected $\bar{\Gamma}_{111}$ established by LEED. SP-UPS spectra independently recover the Bi(111) dispersion and spin reversal in the same local reference frame (see Supplemental Material, Sec.~IV \cite{SupplementalMaterial,Hirahara2007,Shen2019}).

For a nonmagnetic system, time-reversal symmetry imposes opposite spin spectral weight at opposite momenta. At a time-reversal-invariant momentum, the summed spin spectral function vanishes because of Kramers degeneracy. The momentum about which the measured polarization changes sign therefore identifies the crystallographic reference frame followed by the dominant unoccupied response. A response referenced to the macroscopic vicinal surface would reverse at $\bar{\Gamma}_{223}$, corresponding to normal electron incidence, whereas a state referenced to the local Bi(111) terraces must reverse at $\bar{\Gamma}_{111}$. This criterion provides a direct means to compare flat and vicinal Bi films, and spin-integrated intensity alone may not identify this reference frame.

The broad structures $\alpha$, $\beta$, and $\gamma$ correspond qualitatively to the calculated spectral function (Figs.~\ref{fig:Fig4}(a)--\ref{fig:Fig4}(c)), but the approximately 700-meV experimental resolution and transition matrix elements preclude a band-by-band identification from the intensity maxima. Moreover, the maximum of the spin-integrated intensity in Figs.~\ref{fig:Fig4}(a) and \ref{fig:Fig4}(b) does not generally coincide with the maximum of the spin polarization in Figs.~\ref{fig:Fig4}(d) and \ref{fig:Fig4}(e). Motivated by this experimental observation, we use a fitting procedure to locate the low-energy spin-dependent contribution associated with the Bi(111) surface states \cite{Fukushima2024}. The fitted peak positions are superimposed on the experimental data as circles whose radii are proportional to the corresponding peak areas (the raw spectra and fitting analysis are reported in Supplemental Material, Secs.~III and IV \cite{SupplementalMaterial,Hirahara2007,Shen2019}).

Flat Bi/Ge(111) changes sign across $\bar{\Gamma}_{111}$, and the low-energy region \#1 on Bi/Ge(223) reverses about the same point in the local terrace frame (Figs.~\ref{fig:Fig4}(d) and \ref{fig:Fig4}(e)). On flat Bi/Ge(111), region \#2 does not retain the expected antisymmetric behavior, whereas the higher-energy contrast in region \#3 is weaker and less uniform, consistent with averaging over rotational domains.

The KKR calculations provide an initial-state benchmark for interpreting the SP-IPES polarization. The calculation describes a semi-infinite Bi(111) surface rather than the full Bi/Ge(223) interface. Along the $\bar{\Gamma}\bar{M}$ direction, the calculated polarization is dominated by the same transverse in-plane spin component measured experimentally, changes sign at $\bar{\Gamma}$, and qualitatively reproduces the sequence of spin inversions labelled as \#1--\#3 in Fig.~\ref{fig:Fig4}(f). Layer-resolved results show that the corresponding spin-polarized spectral weight is concentrated near the Bi surface (see Supplemental Material, Sec.~VII \cite{SupplementalMaterial,Ebert2011,Nicolai2019}).

 The calculations shown here do not include the one-step corrections required to reproduce the experimental IPES intensities \cite{Wissing2013}. Moreover, the measured spin character can differ from the calculated spin texture of the bands because of experimental geometry and transition matrix elements; spin information is generally more reliable for states of pure orbital symmetry \cite{Wissing2014}, although the orbital character of the present Bi/Ge contributions has not been independently established. Accordingly, the comparison should be restricted to the sign and energy--momentum distribution of the transverse spin response.

Residual rotational replicas on flat Bi/Ge(111) can average over nonequivalent azimuths, whereas Ge(223) selects a dominant registry that is closer to the single-domain surface considered in the KKR calculation. Reduced rotational averaging may therefore explain the clearer agreement observed at higher energies for Bi$(9\,9\,11)$.

 In summary, under identical preparation conditions, different Ge cuts produce different Bi structures and electronic reference frames. The comparison between Ge(443) and Ge(331) suggests that their common Type-II step-edge geometry favors PC-Bi formation despite their different surface reconstructions. The characteristic PC-Bi SP-IPES response follows the macroscopic vicinal plane. Ge(223), instead, yields an ordered Bi$(9\,9\,11)$ vicinal film. Its spin polarization reverses at the local $\bar{\Gamma}_{111}$, in agreement with the spin component and spin-reversal pattern obtained from the calculated unoccupied-state spectral function of Bi(111). In the measured $\bar{\Gamma}\bar{M}$ sector, the initial-state spin-reversal pattern survives the inverse photoemission process. Vicinal Ge(223) therefore provides a semiconductor-supported Bi(111) film whose crystallographic structure and unoccupied spin polarization are referenced to the same local crystal plane.

\begin{acknowledgments}
We thank Dr. I. Majumdar for technical assistance and Prof. A. Crepaldi for useful discussions.
J.M., L.N. and A.P. would like to thank the project QM4ST with Reg. No. CZ.02.01.01/00/22 008/0004572, co-funded by the ERDF as part of the MŠMT and support by the Ministry of Education, Youth and Sports of the Czech Republic through the e-INFRA CZ (ID:90254).
ChatGPT and Codex (OpenAI, primarily GPT-5.5) supported manuscript preparation under author supervision.
\end{acknowledgments}

\paragraph*{Data availability.}
The data that support the findings of this study are available from the corresponding authors upon reasonable request.

\bibliography{Bibliografia_official}

\end{document}